\documentclass[conference]{IEEEtran}
\IEEEoverridecommandlockouts
\usepackage{balance}
\usepackage{comment}
\usepackage{svg}

\usepackage{graphicx}
\usepackage{cite}
\usepackage{amsmath,amssymb,amsfonts}
\usepackage{algorithmicx}
\usepackage{textcomp}
\usepackage{xcolor}

\usepackage{amsthm}

\usepackage[caption=false]{subfig}

\usepackage{authblk}

\usepackage{algorithm,algpseudocode}

\usepackage{multicol}

\usepackage{mathtools}

\usepackage{cuted}

\usepackage{color,soul}
\def\BibTeX{{\rm B\kern-.05em{\sc i\kern-.025em b}\kern-.08em
    T\kern-.1667em\lower.7ex\hbox{E}\kern-.125emX}}

\makeatletter
\newcommand{\linebreakand}{%
\end{@IEEEauthorhalign}
\hfill\mbox{}\par
\mbox{}\hfill\begin{@IEEEauthorhalign}
}

\makeatother

\begin{document}

\title{
Learning to Count Targets from Dual-Window: \\ A CNN Approach for OFDM ISAC

% AKH
\thanks{This work has been supported by the Smart Networks and Services Joint Undertaking (SNS JU) project 6G-DISAC under the EU’s Horizon Europe research and innovation program under Grant Agreement no. 101139130.}
}

\author[1]{Ali Al Khansa}
%\author[1, 2]{Youssef Bahannis}
%\author[2]{Giyyarpuram Madhusudan}
%\author[2]{Guillaume Larue}
%\author[2]{Louis-Adrien Dufrène}

\affil[1]{Orange Labs, Rennes, France}
%\affil[2]{INSA, Rennes, France}

\affil[ ]{ali.alkhansa@orange.com}

\maketitle

\IEEEpubidadjcol

\begin{abstract}

Integrated Sensing and Communication (ISAC) with Orthogonal Frequency Division Multiplexing (OFDM) waveforms is a key enabler for next-generation wireless systems. Recent studies show that Convolutional Neural Networks (CNNs) can estimate the number of targets from two-dimensional (2D) range–Doppler periodogram maps, yet accuracy often degrades as scenes become denser. One significant factor is the classical resolution–sidelobe attenuation trade-off, which limits performance when targets are weak or closely spaced. While windowing is routinely applied to shape this trade-off, the choice is typically static. This paper proposes a new CNN method that uses two windowed range-Doppler periodograms and learns to fuse complementary views: one window optimized for resolution and one window optimized for sidelobe suppression. %The network uses window-specific feature stems followed by a lightweight attention gate to adaptively weight the two views.
The design explicitly targets the resolution–sidelobe attenuation trade-off by exposing the model to complementary windowed maps and letting it learn when each is most informative. Numerical experiments show consistent gains over single-window CNN baselines, with better scaling in target density and greater robustness across different noise levels.

\end{abstract}

\vspace{0.5cm}

\section{Introduction}

\IEEEPARstart{N}{\lowercase{e}}xt-generation wireless networks increasingly seek to unify communication and sensing within a single infrastructure. Accordingly, Integrated Sensing and Communication (ISAC) has gained prominence as a unifying framework for future wireless systems, aiming to co-design waveforms, signal processing, and resource allocation so that a single platform can observe the environment while delivering communication services \cite{strinati2024distributed}. By sharing spectrum, hardware, and timing, ISAC promises tighter situational awareness with reduced cost and latency relative to standalone radar and communication systems. This integration also opens the door to cross-layer optimizations where sensing quality and link performance can be traded according to application needs \cite{zhu2024irs}.

Reflecting current standardization trends, the 3rd Generation Partnership Project (3GPP) \cite{3GPPRAN1} has endorsed an initial 6G physical-layer baseline that retains Cyclic Prefix Orthogonal Frequency Division Multiplexity (CP-OFDM) on the downlink and adopts Discrete Fourier Transform-spread (DFT-s)-OFDM on the uplink, signaling OFDM’s continued centrality in future systems \cite{mcns_6g_agreement_2025}. This choice is motivated by OFDM’s orthogonality, robustness to multipath, flexible resource allocation, and its compatibility with millimeter-wave and massive MIMO deployments \cite{smeenk2024optimizing}. Beyond communications, the multicarrier structure yields sensing-friendly observables making OFDM a natural choice for ISAC \cite{liu2025cp}.

There are a variety of approaches for estimating targets with OFDM sensing \cite[Sec. IV-E]{zhang2021enabling}. At the high-complexity end, some examples are the super-resolution and parametric methods such as MUltiple SIgnal Classification (MUSIC) and the Estimation of Signal Parameters via Rotational Invariance Techniques (ESPRIT) \cite{stoica1997introduction}. Other methods like Bayesian model-based excel under clean priors but typically demand careful tuning and higher compute \cite{jafri2024bayesian}. At the low-complexity end, the periodogram-based methods remain attractive in ISAC systems because they are hardware-friendly, robust across operating conditions, and easy to deploy at scale \cite{mohr2024measurement}. Just as important for this work, the periodogram naturally yields 2-D range–Doppler intensity maps that serve as well-structured inputs for learning: CNNs can ingest these images directly, enabling data-driven detectors to learn range–Doppler patterns leading to target detection and distance and velocity estimation \cite{jeon2024velocity}. 

However, a fundamental trade-off confronts these algorithms: resolution versus sidelobe attenuation. Improving resolution typically increases spectral leakage and elevates sidelobes, whereas suppressing sidelobes broadens the main lobe and degrades the ability to separate close targets \cite{li2024low}. In dense scenes or low Signal to Noise Ratio (SNR) regimes, this trade-off might directly impact peak separability and detection accuracy \cite{choi2021multiple}.% and the challenge intensifies with large amplitude disparities among targets.

Window functions are the standard tool to navigate this trade-off \cite{yu2024rapid}. Resolution-optimized (e.g., rectangular) windows preserve resolution but exhibit pronounced sidelobes, while sidelobe-optimized windows (e.g., Hann, Hamming, Blackman) reduce sidelobes at the cost of main-lobe broadening \cite{braun2014ofdm}. In practice, a single window is usually fixed offline and applied uniformly across frames and scenes, even though the optimal choice depends on the actual scenario (e.g., target density, spacing, and SNR). This motivates designs that adapt to scene conditions or combine complementary window properties rather than committing to a single choice.

After the range–Doppler periodogram is formed, there are different paths for multi-target detection. Classical approaches apply peak detection (often using a Constant False Alarm Rate (CFAR) threshold rules) followed by successive target cancellation procedures to iteratively detect the multiple targets (starting by strong targets, removing their effects, and continuing with weaker ones) \cite{braun2014ofdm}. These approaches are attractive for their simplicity, but their performance can further be improved. Novel approaches uses learning-based detectors, most commonly CNN classifiers that infer the target count from a 2D periodogram map, that localize peaks directly \cite{wei2024multiple,choi2022estimation,jeon2024velocity}. 

Despite their differences, both families usually operate on a single windowed periodogram, and thus both inherit the resolution–sidelobe trade-off discussed earlier. To address this, our prior work proposed dual-windowing as a simple but effective way to expose complementary views of the scene \cite{2WPAPER}. The strategy was to combine the benefits of both resolution-optimized and sidelobe-attenuation-optimized window functions. Specifically, two periodograms were computed using different window functions. Then, a multi-target detection algorithm was applied to each periodogram independently, and the resulting target lists were compared and matched. If the two results aligned, the system adopted the resolution-optimized estimates; if mismatching result was detected, a more complex detection algorithm was triggered to resolve the ambiguity.

While the approach in \cite{2WPAPER} shows improved detection performance and adaptability across diverse scenarios, it also introduces additional computational steps. As the number of targets increases, the complexity of the matching process becomes non-negligible. Moreover, invoking more complex cancellation algorithms when ambiguities are detected adds further processing overhead, making this method less suitable in many scenarios like applications with a high density of targets.

Motivated by these limits, the present paper keeps the dual-window principle but replaces explicit matching with learning-based CNN methods. Specifically, we propose feeding the two windowed periodograms as complementary channels to a CNN that is trained end-to-end to infer the number of targets. This eliminates the run-time pairing/search stage and lets the model learn when each window is informative. Because training is offline, we can afford richer backbones or fusion modules without incurring inference-time complexity comparable to combinatorial matching as seen in \cite{2WPAPER}.

Recent CNN-based estimators for OFDM radar demonstrate that learning from a single windowed periodogram can yield competitive count estimates, but performance typically degrades as the number of targets increases and as sidelobe contamination rises \cite{wei2024multiple,choi2022estimation,jeon2024velocity}. In essence, the underlying single-view representation still embodies the same resolution–sidelobe compromise, which becomes the bottleneck at higher densities or lower SNRs.

Our goal is to outperform both the dual-window matching baseline \cite{2WPAPER} (by removing its complexity and matching fragility) and single-window CNN baselines \cite{wei2024multiple,choi2022estimation,jeon2024velocity} (by giving the network complementary views and learning the fusion). We will show in simulation that dual-window CNNs deliver higher accuracy, more graceful scaling with target density, and stronger robustness across SNRs than their single-window counterparts, while avoiding the combinatorial overhead of explicit matching.

The rest of the paper is organized as follows: Section II presents the system model and signal processing pipeline for generating dual-window periodograms. Section III details the proposed dual-window CNN architectures and the training protocol. Section IV reports numerical results and findings. Section V concludes and outlines future directions.

\section{OFDM System Model and  Periodogram Calculations}

As our CNN will be trained on dual range–Doppler periodograms, we first detail how these maps are formed. This section presents the OFDM signal model used to obtain the time–frequency frame, then derives the 2D periodogram that yields the range–Doppler map, and finally explains where and how window functions are concerned, and how they shape resolution and sidelobe behavior.

\subsection{OFDM Radar Fundamentals}
In OFDM radar, the received waveform $r(t)$ is the sum of echoes from $H$ scatterers, each contributing with its own attenuation, propagation delay, and Doppler shift. A convenient model is therefore \cite{braun2014ofdm}:
\begin{equation}
    r(t) = \sum_{h=0}^{H-1} b_h s(t-\tau_h) e^{j 2 \pi f_{D,h} t} e^{j \overline{\phi}_h} + \overline{z}(t),
\end{equation}
where 
\begin{itemize}
    \item $b_h$ is the attenuation factor, 
    \item $\tau_h$ represents the round-trip delay, 
    \item $f_{D,h}$ represents the Doppler frequency shift, 
    \item $\overline{\phi}_h$ is a random phase offset, 
    $\overline{z}(t)$ denotes Additive White Gaussian Noise (AWGN).
\end{itemize}
The transmitted signal $s(t)$, as defined in the OFDM scheme, comprises subcarriers that are orthogonal in time and frequency, ensuring efficient spectral utilization and minimizing interference.

Applying the previous equation to OFDM signals and introducing a new notation, a transmitted OFDM frame can be written as:
\begin{equation}
\label{F_Tx}
    \textbf{\textit{F}}_{T_x} = \begin{pmatrix}
c_{0,0} & \cdots & c_{0,M-1} \\
\vdots & \ddots & \vdots \\
c_{N-1,0} & \cdots & c_{N-1,M-1}
\end{pmatrix},
\end{equation}
where
\begin{itemize}
    \item symbols $c_{k,l}$, $k\in\{0,\dots,N-1\}$ and $l\in\{0,\dots,M-1\}$ are chosen from a modulation alphabet (e.g., QPSK, QAM, etc.), 
    \item $N$ is the number of subcarriers, 
    \item $M$ is the number of OFDM symbols. 
\end{itemize}
A row in $\textbf{\textit{F}}_{T_x}$ represents a subcarrier and a column represents an OFDM symbol. After an analog-to-digital conversion, followed by demodulation, the received frame matrix $\textbf{\textit{F}}_{R_x}$, captures the effects of the propagation channel. For $H=1$, i.e., in case we have a single target, the received signal can be written as:
\small \begin{equation}
\label{F_Rx}
(\textbf{\textit{F}}_{R_x})_{k,l} = b_0(\textbf{\textit{F}}_{T_x})_{k,l} e^{j 2 \pi T_O f_{D,0}l} e^{-j 2 \pi \tau_0(k \Delta f + f_0)} e^{j \overline{\phi}_0} + (\overline{\textbf{\textit{Z}}})_{k,l},
\end{equation} \normalsize 
where \begin{itemize}
    \item $\overline{\textbf{\textit{Z}}} \in \mathbb{C}^{N \times M}$ is the matrix representation of the AWGN,
    \item $f_0$ is the initial frequency of the $N$ subcarriers (i.e., the frequencies go from $f_0$ through $f_{N-1}$), 
    \item $T_O$ is the OFDM symbol duration (with the CP duration),
    \item $\Delta f$ is the subcarrier spacing.
\end{itemize} 
As $f_0$ and $\overline{\phi}_0$ are constant, define $\phi_h=\overline{\phi}_h-2\pi f_0 \tau_h$. Element-wise division of the received matrix by the transmitted matrix is performed to isolate the effects of target reflections, producing the normalized frame matrix \textbf{\textit{F}}, expressed as: 
\begin{equation}
\label{F_0}
    (\textbf{\textit{F}})_{k,l} = b_0 e^{j 2 \pi l T_O f_{D,0}} e^{-j 2 \pi k \tau_0 \Delta f} e^{j \phi_0} + (\textbf{\textit{Z}})_{k,l},
\end{equation}
where 
$(\textbf{\textit{Z}})_{k,l}= (\overline{\textbf{\textit{Z}}})_{k,l} / \textbf{\textit{F}}_{T_x}$ represents the normalized AWGN. From here, the estimation problem is seen as a spectral estimation problem, where the delay $\tau$ and the Doppler frequency $f_D$ (parameters of interest) correspond to the target's distance and relative velocity, respectively.

Since the operations used to calculate $\textbf{\textit{F}}_{R_x}$ from $r(t)$ are linear with respect to their input signal, the result can be generalized to $H>1$ targets:
\begin{equation}
\label{F_general}
    (\textbf{\textit{F}})_{k,l} = \sum_{h=0}^{H-1} b_h e^{j 2 \pi l T_O f_{D,h}} e^{-j 2 \pi k \tau_h \Delta f} e^{j \phi_h} + (\textbf{\textit{Z}})_{k,l}.
\end{equation}

\subsection{Periodogram-Based Estimation}
Next, we present the formulation of periodogram, and show how it can be used for target detection and parameter estimation. Both cases of conventional methods and CNN methods are to be illustrated. We will show how the periodogram is used as a spectral estimation tool to estimate the target parameters from $\textbf{\textit{F}}$. As defined in \cite{braun2014ofdm} Eq. (3.30), the two-dimensional periodogram for the received data is defined as:
\begin{equation}
\label{Per_F}
    \text{Per}_\textbf{\textit{F}}(n, m) = \frac{1}{NM} \left| \sum_{k=0}^{N_{\text{Per}}-1} \sum_{l=0}^{M_{\text{Per}}-1} (\textbf{\textit{F}})_{k,l} e^{-j 2 \pi \frac{l m}{M_{\text{Per}}}} e^{j 2 \pi \frac{k n}{N_{\text{Per}}}} \right|^2,
\end{equation}
where 
$N_{\text{Per}}$ is the size of the Inverse Fast Fourier Transforms (IFFT) and $M_{\text{Per}}$ is the size of the Fast Fourier Transforms (FFT). These transforms are applied to the rows and columns of \textbf{\textit{F}}, with the aim of isolating the sinusoidal components corresponding to the targets' Doppler shifts and delays. The peaks (local maxima) in the periodogram represent the estimated parameters, and can be translated into target distances and velocities.

%Fig. \ref{fig:3.6} illustrates a block diagram of a periodogram-based OFDM ISAC system. The diagram begins with a monostatic ISAC transceiver that transmits a signal to a communication user while simultaneously receiving reflected signals from a target. The signal processing chain starts by dividing the received signal by the transmitted signal, as described in Eq. (\ref{F_general}). Following this, the periodogram is calculated to generate the range–Doppler map. From this map, two branches are possible: on the right, conventional processing applies peak-based detection to estimate the target count; on the left, learning-based approaches feed the range–Doppler map to a CNN that is trained to predict the number of targets directly.

\subsection{Windowing for Resolution-Sidelobe Trade-off}

\begin{figure}
\centering
\includegraphics[scale=0.48]{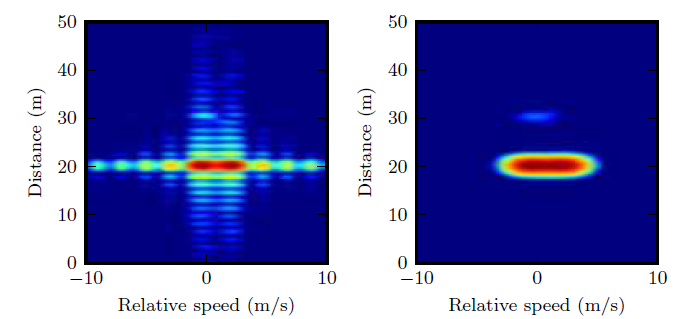}
            \caption{Periodogram of three targets using two different windows: (a) rectangular window optimized for resolution: the two targets at 20m are easily distinguished, (b) Dolph-Chebyshev window optimized for sidelobe attenuation: the target at 30m now clearly stands out \cite{braun2014ofdm}.}
\label{fig:3.10'}
\end{figure}

Once the range–Doppler periodogram is formed, a windowing function is typically applied to steer the trade-off between resolving closely spaced peaks and suppressing leakage from strong scatterers. In short, high range resolution helps separate close targets, whereas strong sidelobe attenuation prevents weak echoes from being masked by the sidelobes of dominant returns. To optimize this trade-off, designs must choose an operating point that reflects scene conditions and requirements.

Windowing achieves this by pre-shaping the data prior to the 2D transforms (i.e., within the periodogram), thereby modifying the spectrum of the resulting map. A rectangular window preserves a narrow mainlobe and is often taken as a resolution benchmark, but it exhibits relatively high sidelobes. In contrast, sidelobe-suppressed windows such as Hann, Hamming or Blackman families reduce sidelobe levels substantially at the cost of broadening the mainlobe and thus degrading separability of nearby peaks. More specialized designs (e.g., Dolph–Chebyshev) allow explicit control over sidelobe levels with a corresponding mainlobe penalty\cite{braun2014ofdm}.

To incorporate a windowing in the periodogram, a two-dimensional window matrix $\textbf{\textit{W}}$ is defined, which is multiplied element-wise with the matrix $\textbf{\textit{F}}$. Accordingly, equation (\ref{Per_F}) can be rewritten as:
\begin{align}
    \label{Per_F_Window}
    & \text{Per}_\textbf{\textit{F}}(n, m) =  \\ \nonumber
    & \frac{1}{NM} \left| \sum_{k=0}^{N_{\text{Per}}-1} \sum_{l=0}^{M_{\text{Per}}-1} (\textbf{\textit{F}})_{k,l}(\textbf{\textit{W}})_{k,l} e^{-j 2 \pi \frac{l m}{M_{\text{Per}}}} e^{j 2 \pi \frac{k n}{N_{\text{Per}}}} \right|^2.
\end{align}
Fig. \ref{fig:3.10'} qualitatively contrasts two cases using identical scenes: a rectangular window yielding a narrow mainlobe and better separation of close targets, and a stronger sidelobe windowing yielding lower sidelobes and improved visibility of weak targets near stronger ones, but with reduced resolution. This inherent trade-off motivates methods that can adapt the effective tapering to the scene or combine complementary properties, rather than committing to a single fixed window for all conditions. The contribution of this paper lies in leveraging a dual-window approach that simultaneously considers both resolution and sidelobe attenuation, thereby balancing detection performance across diverse conditions. The idea will be implemented in a CNN training, aiming to outperform the performance of single-window CNN methods seen in the State-of-the-Art (SotA).

\section{The Dual-Window CNN Model}

This section first reviews different methods that operate on range–Doppler periodograms, both classical and learning-based, and emphasizes their shared reliance on a single window. We then introduce our dual-window CNN, which learns to fuse complementary periodogram views in order to mitigate the resolution–sidelobe trade-off.
\subsection{Conventional Methods}

\begin{figure}[t]
\vspace{20 pt}
\centering
\includegraphics[scale=0.23]{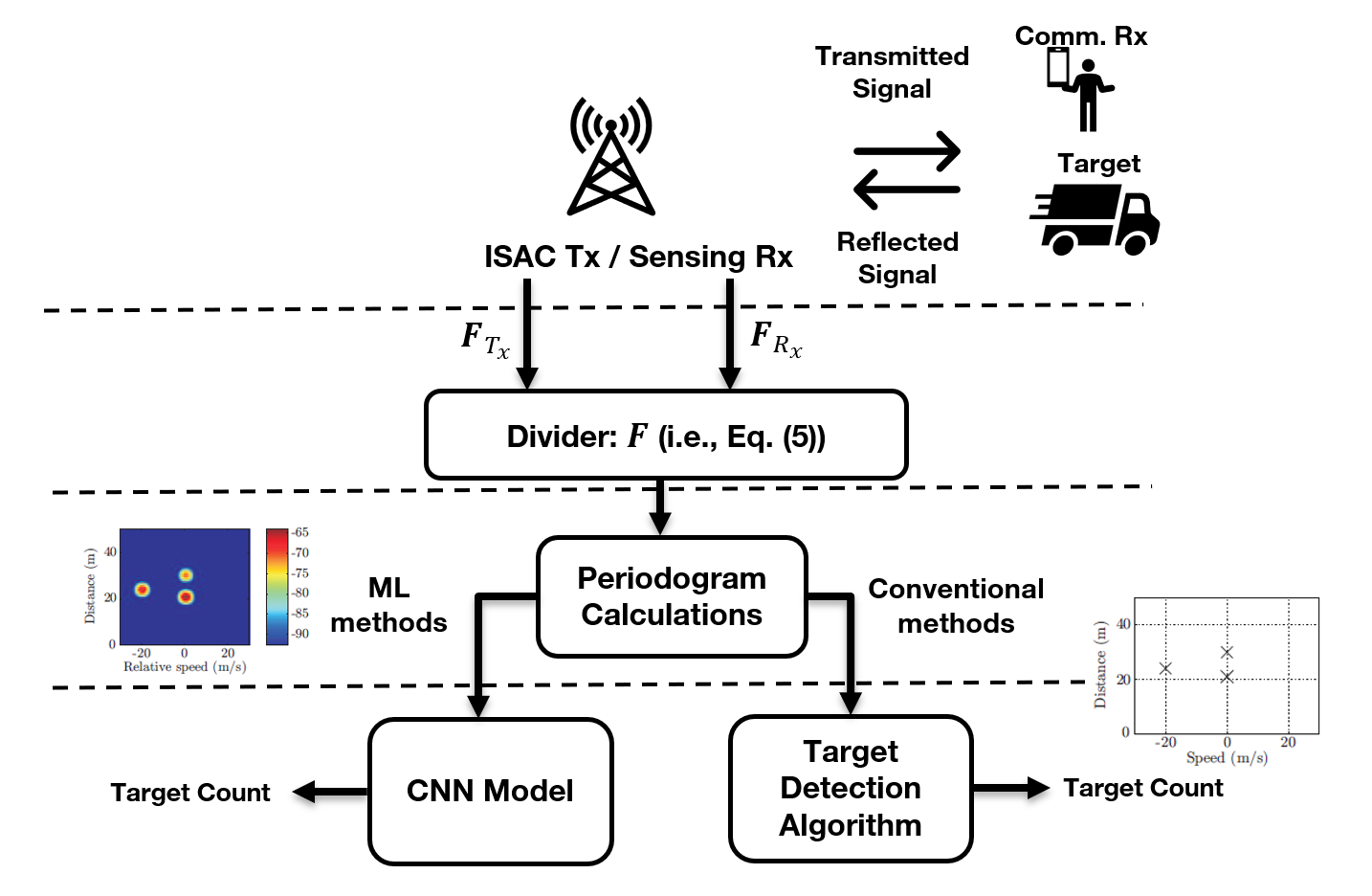}
\caption{Block diagram of a periodogram-based OFDM ISAC system. Right side showing conventional methods; left side showing CNN models.} %The contributions of this paper relate to the highlighted yellow parts: the windowing and the target detection algorithm.}
\label{fig:3.6}
\end{figure}

In Fig. \ref{fig:3.6}, we sketch a periodogram-based OFDM ISAC chain. A monostatic transceiver emits an OFDM frame toward a communication user while receiving echoes from surrounding scatterers. After downconversion and demodulation, the received frame is divided element-wise by the transmitted frame to form the normalized symbol matrix as described in Eq. (\ref{F_general}). Then, the periodogram is calculated as in Eq. (\ref{Per_F_Window}) to generate the range–Doppler map. From this map, two branches are possible: on the right, conventional processing applies peak detection (often with CFAR-like rules) followed by iterative cancellation algorithms \cite{braun2014ofdm} to estimate the target count; on the left, learning-based approaches feed the range–Doppler map to a CNN \cite{wei2024multiple,choi2022estimation,jeon2024velocity} that is trained to predict the number of targets directly.

%Our recent work \cite{2WPAPER} extends this idea by generating a resolution-oriented map and a sidelobe-suppressed map, then reconciling the two through a matching stage to control complexity while preserving accuracy. In practice, however, as target density increases, the matching stage in \cite{2WPAPER} becomes the dominant cost and an accuracy bottleneck.

Learning-based methods replace hand-crafted postprocessing with a CNN that ingests a windowed periodogram and outputs the target count as a multi-class label. This yields simple inference and avoids iterative cancellation, but the input representation still relies on a single window. As scenes become dense or exhibit large dynamic range, a fixed taper cannot jointly provide fine resolution and strong sidelobe suppression, and accuracy degrades accordingly. These limitations motivate a learned fusion of complementary windowed views rather than committing to a single choice.

\subsection{Proposed Dual-Window Method}

We propose a learning-based detector that uses two windowed range–Doppler periodograms computed from the same noisy echo and learns to fuse their complementary information. Concretely, we use two 2D windows, one oriented toward resolution and another toward sidelobe suppression, and stack the resulting maps as two aligned views of the same scene. The network predicts the target count $K \in \{1,\dots,H_t\}$ via an $H_t$-way softmax trained with cross-entropy, with $H_t$ representing the maximum class. The remainder of this section details how the dual-window model is designed, trained, and evaluated.

A variety of CNN architectures can be adapted for target-count estimation. Following the spirit of the SotA (e.g., \cite{choi2021multiple}), multiple convolutional layers extract spatial features from the periodogram maps, followed by a classifier head that outputs the final prediction. While alternative backbones could be used, the key design element here is the use of complementary, windowed inputs, and a classifier that operates directly on these inputs to estimate the number of targets. In the dual-window case, the architecture is adapted to handle a two-channel input, with each channel corresponding to one of the two periodograms.

\begin{figure}
\vspace{40 pt}
\centering
\includegraphics[scale=0.21]{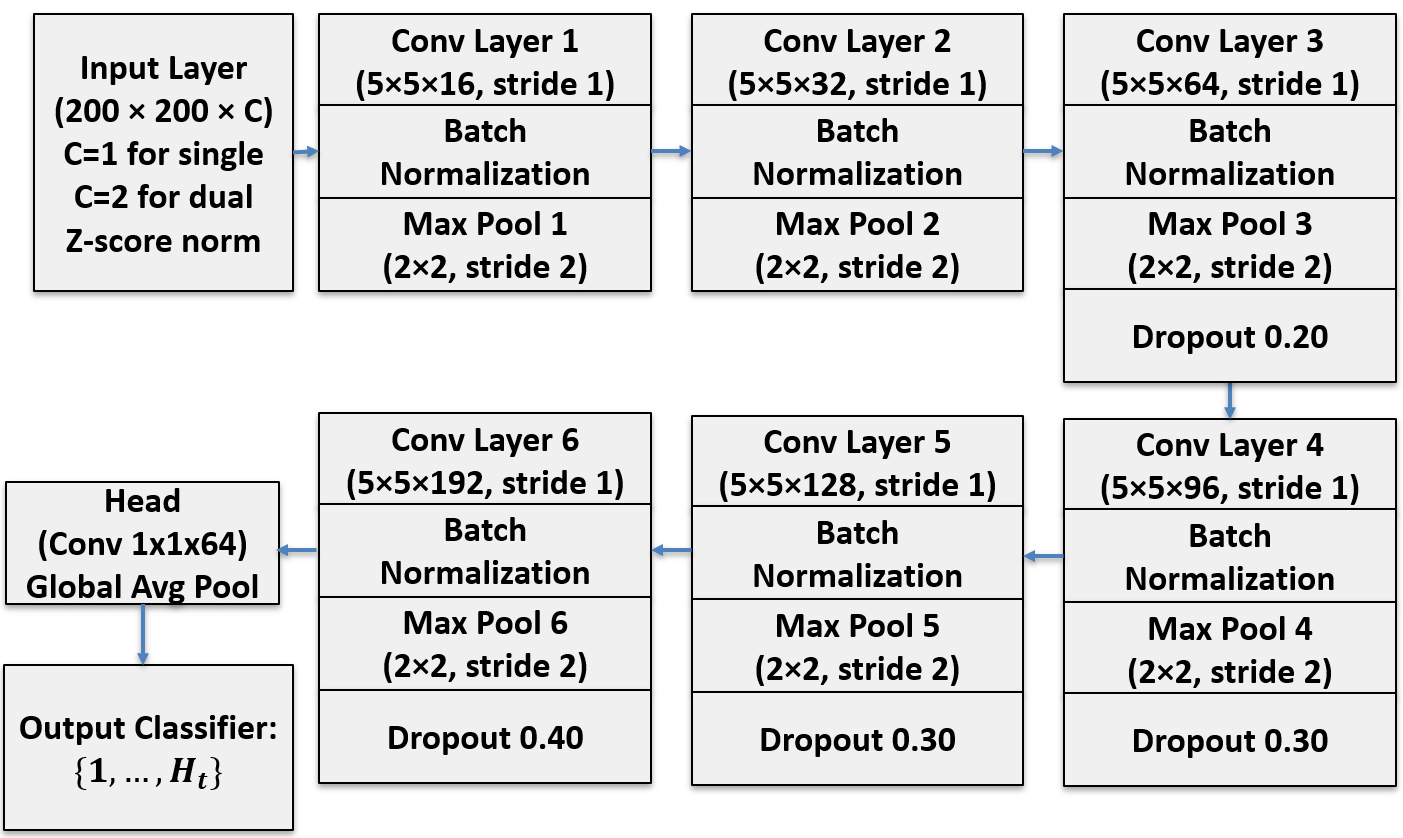}
            \caption{The proposed CNN architecture.}
\label{CNNMODEL}
\end{figure}

Specifically, our classifier employs six $5{\times}5$ convolutional blocks with channel widths $\{16, 32, 64, 96, 128, 192\}$. Each block follows the sequence:
\[
\text{Conv}(5{\times}5,\,C)\ \rightarrow\ \text{BatchNorm}\ \rightarrow\ \newline \text{ReLU}\ \rightarrow\ \text{MaxPool}.
\]
To improve generalization, dropout is inserted after blocks 3–6 with rates $\{0.20, 0.30, 0.30, 0.40\}$, respectively. After the six blocks, we apply a $1{\times}1$ projection (Conv $1{\times}1$, 64) followed by ReLU and Global Average Pooling (GAP), then a fully connected layer to $H_t$ classes and a softmax. The input layer is $200{\times}200{\times}C$ with z-score normalization ($C{=}1$ for the single-window baseline; $C{=}2$ for the dual-window early-fusion setup where the rectangular and the Hann periodograms are stacked channel-wise). Fig. \ref{CNNMODEL} shows a block diagram of the proposed two-window CNN.

%For context, \cite{choi2022estimation} evaluates two additional CNN baselines (a 7-layer and a 4-layer model), reporting higher accuracy for the deeper network at increased complexity. A similar work uses YOLO architecture is seen in \cite{jeon2024velocity}. These approaches degrade as the number of targets grows and neither explicitly addresses the resolution–sidelobe attenuation trade-off. Our dual-window formulation targets this gap by providing complementary windowed inputs to the CNN, enabling robustness as scene density increases.

%We study two ways to fuse the two windows views. The first way is the early fusion, and which corresponds to the illustration in Fig. \ref{CNNMODEL}. This keeps the parameter budget nearly unchanged and lets the first layer learn filters jointly over both views, but it forces that layer to handle both window-specific feature extraction and inter-window mixing simultaneously. Another way is to split stems with lightweight gating. In this case, each window first passes through a short, independent stem of two convolutional blocks (Conv–BN–ReLU–Pool) that learn window-specific primitives. The outputs are depth-concatenated and fused with a $1\times1$ bottleneck. A compact squeeze-and-excitation gate then produces channel-wise weights from global-average pooled context and reweights the fused tensor before a shared trunk. This reduces interference between views and enables scene-adaptive emphasis: in sparse scenes the model can lean on the resolution-oriented map, while in dense or high-dynamic-range scenes it can upweight the sidelobe-suppressed map. After fusion, both dual variants use the same trunk as the baselines: several Conv–BN–ReLU blocks with pooling and dropout, a $1\times1$ projection, global-average pooling, and a fully connected classifier over $H_t$ classes. Global-average pooling controls parameter count and improves robustness to small spatial shifts. \rv{This methods shows better gains, and thus will be adapted in the numerical results section}.

\section{Numerical Results}
\subsection*{Setup and Training}
In this section, we validate our proposal via simulations that are implemented in \textsc{Matlab} (Deep Learning Toolbox). We consider a monostatic OFDM-ISAC transceiver using CP-OFDM frames with $N_{\text{use}}=1284$ active subcarriers out of $N = N_{\text{Per}} = 4096$ and subcarrier spacing $\Delta f{=}30$\,kHz (occupied bandwidth $B=N_{\text{use}}\Delta f$). In the simulations, $M$ is fixed to 64 and $M_{\text{Per}}$ to 256 \cite{choi2021multiple}. A cyclic prefix of duration $T_{\text{CP}}$ is appended to each OFDM symbol and chosen long enough to exceed the maximum two-way delay spread; CP is removed prior to demodulation and does not affect the periodogram geometry. The carrier frequency is $28$\,GHz. %Slow-time sampling uses a pulse repetition interval $\text{PRI}=10^{-4}$\,s with $M_{\text{FFT}}{=}256$ pulses per coherent processing interval (CPI). We form range–Doppler periodograms on an $N_{\text{FFT}}{\times}M_{\text{FFT}}$ grid with $N_{\text{FFT}}{=}4096$ and center-crop to $200{\times}200$ pixels for learning. The resulting physical bin spacings are
%\[
%d_R \;=\; \frac{c_0}{2B}\,\frac{N_{\text{use}}}{N_{\text{FFT}}}\quad\text{meters/bin},\qquad
%d_v \;=\; \frac{\lambda}{2 M_{\text{FFT}}\!\cdot\!\text{PRI}}\,\text{m/s per bin}.
%\]

Targets are placed at fractional-bin locations inside the $200{\times}200$ range-Doppler crops. The complex echo of $K$ point targets is generated on the slow–fast time lattice with random phases and amplitudes following a two-way path-loss model, normalized at a reference range within the field of view. Additive white Gaussian noise is added such that the noise variance is set from the average per-target echo power so that the time-domain echo-to-noise ratio equals the requested SNR. From the same noisy echo, we form one or two periodograms by applying separable windows in range and Doppler (Rectangular for resolution, Hann for sidelobe suppression), and we input them to the CNN model detailed in the previous section. For the dual-window input, the two periodograms are stacked as channels of a single $200{\times}200{\times}2$ tensor. Note that any other pair of windows can be chosen without losing generality, conditioned that the first window is optimized for resolution and the second for sidelobe attenuation.

We train classifiers to predict the target count $K\in\{1,\dots,H_t\}$ with $H_t{=}12$. Each training example draws $K$ uniformly from $\{1,\dots,12\}$ and an SNR uniformly from $[-30,9]$\,dB. The training set contains $50{,}000$ samples and the validation set $5{,}000$ samples. Inputs use z-score normalization at the network front end. Optimization uses Adam with an initial learning rate of $10^{-3}$ and standard $\ell_2$ regularization, mini-batch size $100$, and a total of $40$ epochs. %Unless stated otherwise, metrics are reported on Monte Carlo trials generated with the same physical setup and image formation pipeline described above.

%Training uses $H_t{=}12$ classes (target counts $K\in\{1,\dots,12\}$). For each training example, we sample $K$ uniformly from $\{1,\dots,12\}$ and draw the SNR uniformly from $[-30,9]$\,dB. We generate $50{,}000$ training and $5{,}000$ validation samples. Images are z-score normalized at the input. Optimization uses Adam with the same hyperparameters as in the baseline CNN study (initial learning rate $10^{-3}$ and standard $\ell_2$ regularization), a mini-batch size of $100$, and $40$ total epochs. Validation is performed periodically during training, and final models are evaluated on Monte Carlo trials at fixed SNRs and across SNR sweeps, as detailed in the subsequent subsections.

\subsection*{Results}

\begin{figure}
\centering
\includegraphics[scale=0.58]{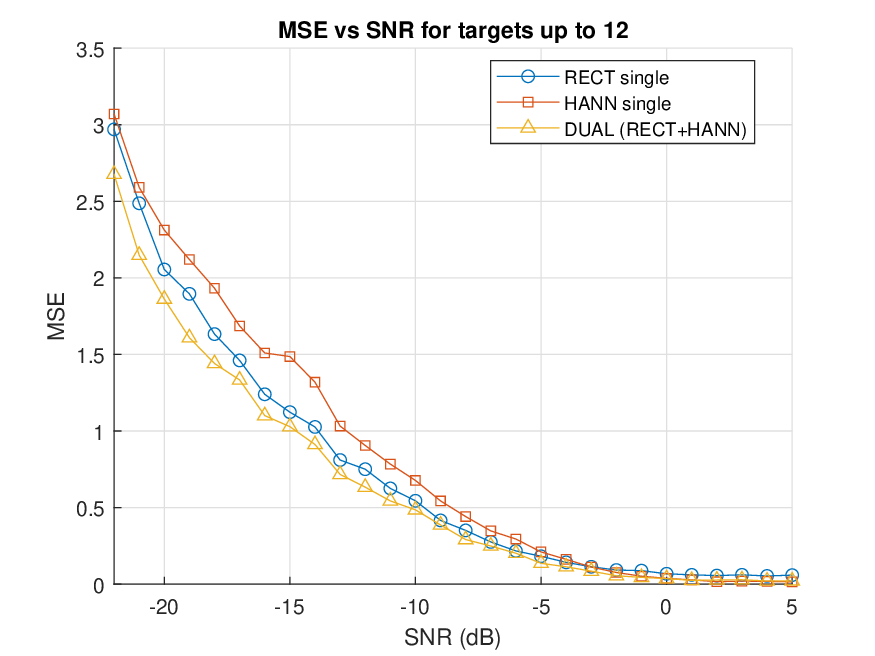}
\caption{Loss Vs SNR}
\label{NR:Ht12_LossvsSNR}
\end{figure}

\begin{figure}
\centering
\includegraphics[scale=0.58]{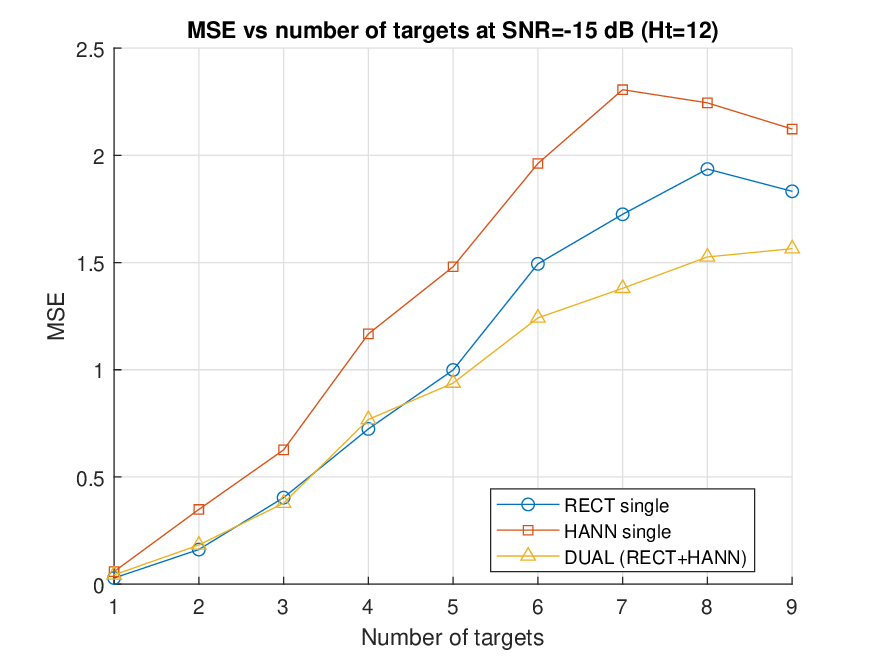}
\caption{Loss Vs Targets}
\label{NR:Ht12_LossvsTargets}
\end{figure}

We compare the performance of the proposed dual-window CNN against two single-window baselines: (i) a resolution-oriented model that uses a rectangular window, and (ii) a sidelobe-suppressed model that uses a Hann window. Several detection metrics could be examined (e.g., perfect-count rate, true/false positives), but in this study we evaluate the counting loss as a the Mean Squared Error (MSE) on the estimated target number. For each Monte Carlo (MC) trial with ground-truth $K$ and estimate $\widehat{K}$, we compute $(\widehat{K}-K)^2$; the reported loss is the MC average. This choice penalizes over-counts and under-counts symmetrically and rewards estimates that are closer to $K$ even when not exact.

Two operating sweeps are considered. First, in Fig. \ref{NR:Ht12_LossvsSNR} we consider the MSE versus the SNR when $K$ is drawn uniformly from $\{1,\dots,H_t\}$ with $H_t{=}12$. Second, in Fig. \ref{NR:Ht12_LossvsTargets}, we present the MSE versus the number of targets at a fixed SNR of $-15$\,dB. 

Considering Fig. \ref{NR:Ht12_LossvsSNR}, we see that the MSE decreases monotonically with SNR for all methods. Across the SNR range, the dual-window model yields consistently lower loss than both single-window baselines. At low SNR, the gap is most visible, reflecting improved robustness when echoes are weak and sidelobes/noise are prominent. As SNR increases, all curves approach near-zero loss and the margin narrows, as the task becomes easier irrespective of window choice. 

Considering Fig. \ref{NR:Ht12_LossvsTargets}, we see that for all methods, the loss grows with scene density (larger $K$) possibly due to increased peak overlap and masking. The dual-window model maintains the lowest loss, with a clearer advantage in the mid-to-high target-count regime. 

These results indicate that exposing the network to complementary windowed views can improve count estimation over a broad range of SNRs and target densities. The runtime overhead is minimal: the two periodograms are formed from the same noisy echo, and inference proceeds on a two-channel input through a single CNN. Training is performed offline, so the additional learning burden does not affect deployment-time latency. Overall, the dual-window formulation offers a simple and effective way to mitigate the resolution–sidelobe tension inherited by single-window pipelines.

\subsection*{Limitations and Future Work}
The presented results show consistent gains of the dual-window CNN over single-window baselines across SNRs and target densities; nevertheless, a broader analysis is warranted. First, this study did not perform an exhaustive architecture search. We focused on a reasonable CNN backbone and two practical fusion designs, but the space of alternatives (e.g., deeper/shallower backbones, different kernel widths, alternative normalization/activation choices, attention-based or cross-window fusion, late vs.\ hybrid fusion) remains largely unexplored. A natural next step is a systematic architecture ablation to identify the most effective design for dual-window inputs, including the choice of fusion points of the windows.

Second, while we trained and evaluated on synthetic OFDM ISAC data with controlled physics and noise, robustness to additional impairments (e.g., hardware nonidealities, clutter, interference, carrier offsets) and transfer to measured datasets require further validation. Future work should therefore include real-data experiments and domain-adaptation strategies.

Finally, the present work targets target-count estimation. Extending the dual-window learning to joint counting and localization (range–Doppler peak attribution), as well as to confidence (uncertainty) estimates, is an important direction for end-to-end ISAC perception.

%Beyond the early-fusion baseline, we also test a lightweight dual-branch variant designed to better exploit the complementarity between the two windowed inputs when the target count range increases to $H_t{=}20$. Each window (Rect and Hann) is first processed by a short, independent stem (two Conv–BN–ReLU–MaxPool blocks per branch). The resulting feature maps are depth-concatenated and passed through a $1{\times}1$ fusion convolution. A shallow squeeze-like gate (a $1{\times}1$ conv followed by a sigmoid) produces per-location weights that modulate the fused features via elementwise multiplication. The gated features then enter the same shared trunk used in the baseline (six 5$\times$5 Conv–BN–ReLU–MaxPool blocks with the same channel widths and dropout placements), followed by a $1{\times}1$ projection, global average pooling, and a fully connected layer to $H_t$ classes.

%We keep the training protocol (optimizer, batch size, epoch budget, data generator, and SNR sampling) consistent, except for adjusting the classifier head to $H_t{=}20$. Figure~\ref{fig:acc_snr_ht20_gated} reports perfect-detection accuracy versus SNR for this variant. The curve shows that the gated fusion maintains a consistent margin over single-window baselines across SNRs while handling the expanded class set, indicating that modest fusion capacity can help the model leverage complementary window information in denser scenes without substantially increasing inference complexity.

\balance
\section{Conclusions}
In this paper, we presented a dual-window learning approach for OFDM ISAC that tackles the resolution–sidelobe trade-off from a data-driven perspective. Instead of committing to a single window, our method feeds two complementary windowed range–Doppler periodograms to a CNN that learns how to detect targets more effectively. Compared with single-window CNN baselines, our method shows an improved performance, more graceful scaling with target count, and stronger robustness across SNRs, while avoiding the matching overhead and sensitivity of some prior arts. Because training occurs offline, inference remains lightweight and compatible with real-time operation.

%Beyond count estimation, the same dual-view design is readily extensible to joint range–Doppler localization and other downstream tasks. Future work will explore richer fusion blocks, and other CNN models, while extending our target counter detector into  curriculum and hard-scene training, and additional complementary tapers or learned windows, as well as transfer to measured data and hardware-in-the-loop evaluation.

\bibliography{conf1}

@article{liu2025cp,
  title={CP-OFDM achieves the lowest average ranging sidelobe under QAM/PSK constellations},
  author={Liu, Fan and Zhang, Ying and Xiong, Yifeng and Li, Shuangyang and Yuan, Weijie and Gao, Feifei and Jin, Shi and Caire, Giuseppe},
  journal={IEEE Transactions on Information Theory},
  year={2025},
  publisher={IEEE}
}

@article{zhang2021enabling,
  title={Enabling joint communication and radar sensing in mobile networks—A survey},
  author={Zhang, J Andrew and Rahman, Md Lushanur and Wu, Kai and Huang, Xiaojing and Guo, Y Jay and Chen, Shanzhi and Yuan, Jinhong},
  journal={IEEE Communications Surveys \& Tutorials},
  volume={24},
  number={1},
  pages={306--345},
  year={2021},
  publisher={IEEE}
}

@INPROCEEDINGS{2WPAPER,

  author={Khansa, Ali Al and Bahannis, Youssef},

  booktitle={IEEE CommNet}, 

  title={Adaptive Dual-Windowing Strategies for Multi-Target Detection in {OFDM ISAC}}, 

  year={2025 },

  volume={},

  number={},

  pages={},
}

@inproceedings{jeon2024velocity,
  title={Velocity and Distance Estimation of Multiple Objects Using {YOLO}-Based {OFDM} Radar System},
  author={Jeon, So-Yeon and Hyun, In-Young and Jeong, Eui-Rim},
  booktitle={2024 15th International Conference on Information and Communication Technology Convergence (ICTC)},
  pages={2214--2216},
  year={2024},
  organization={IEEE}
}

@article{choi2022estimation,
  title={Estimation of number of targets based on CNN classifier for {OFDM} radar systems},
  author={Choi, Jae-Woong and Oh, Jeong-Eun and Jo, A-Min and Jeong, Eui-Rim},
  journal={Mathematical Statistician and Engineering Applications},
  volume={71},
  number={3},
  pages={555--564},
  year={2022}
}

@article{choi2021multiple,
  title={Multiple Target Detection For {OFDM} Radar Based On Convolutional Neural Network},
  author={Choi, Jae-Woong and Jeong, Eui-Rim},
  journal={Turkish Journal of Computer and Mathematics Education},
  volume={12},
  number={6},
  pages={544--550},
  year={2021},
  publisher={Ninety Nine Publication}
}

@inproceedings{jafri2024bayesian,
  title={Bayesian Learning for Sparse Parameter Estimation in OTFS-aided mmWave MIMO Radar Systems},
  author={Jafri, Meesam and Srivastava, Suraj and Jagannatham, Aditya K},
  booktitle={2024 Joint European Conference on Networks and Communications \& 6G Summit (EuCNC/6G Summit)},
  pages={422--427},
  year={2024},
  organization={IEEE}
}

@online{3GPPRAN1,
  title   = { The 3rd {G}eneration {P}artnership {P}roject {R}adio {A}ccess {N}etworks (Physical layer)},
  author  = {{3GPP RAN1}},
  
  date    = {2025-08-29},
  url     = {https://www.3gpp.org/3gpp-groups/radio-access-networks-ran/
},
}

@online{mcns_6g_agreement_2025,
  title   = {First {6G} Agreement Reached: {CP-OFDM} and {DFT-s-OFDM} Confirmed - but is this really the future of {6G?}},
  author  = {{MCNS}},
  
  date    = {2025-08-29},
  url     = {https://mcns5g.com/first-6g-agreement-reached-cp-ofdm-and-dft-s-ofdm-confirmed-but-is-this-really-the-future-of-6g/},
}

@inproceedings{zhu2024irs,
  title={{IRS}-Assisted Collaborative Tracking for {ISAC} in Vehicular Networks},
  author={Zhu, Yuncan and Li, Meiling and El Bouanani, Faissal and Zhao, Yubin},
  booktitle={2024 7th International Conference on Advanced Communication Technologies and Networking (CommNet)},
  pages={1--6},
  year={2024},
  organization={IEEE}
}

@article{stoica1997introduction,
  title={Introduction to spectral analysis},
  author={Stoica, Petre Gheorghe and Moses, Randolph L},
  journal={Prentice Hall},
  year={1997}
}

@phdthesis{mohr2024measurement,
  title={Measurement-Based Performance Analysis of RADAR Estimation Algorithms},
  author={Mohr, Lorenz},
  year={2024},
  school={Universit{\"a}tsbibliothek}
}

@article{yu2024rapid,
  title={A Rapid Single-view Radar Imaging Method with Window Functions},
  author={Yu, Wen Ming and Yang, Yi Ting and Lu, Xiao Fei and Yang, Chao and Chen, Zai Gao and Cui, Tie Jun},
  journal={ACES},
  pages={1--8},
  year={2024}
}

@inproceedings{li2024low,
  title={Low-range-sidelobe waveform design for {MIMO-OFDM ISAC} systems},
  author={Li, Peishi and Xiao, Zichao and Li, Ming and Liu, Rang and Liu, Qian},
  booktitle={ICC 2024-IEEE International Conference on Communications},
  pages={909--914},
  year={2024},
  organization={IEEE}
}

@inproceedings{smeenk2024optimizing,
  title={Optimizing Radio Resources for Radar Services in {ISAC} Systems by Deep Reinforcement Learning},
  author={Smeenk, Carsten and Zhao, Zhixiang and Schneider, Christian and Robert, Joerg and Del Galdo, Giovanni},
  booktitle={2024 IEEE 35th International Symposium on Personal, Indoor and Mobile Radio Communications (PIMRC)},
  pages={1--6},
  year={2024},
  organization={IEEE}
}

@article{wei2024multiple,
  title={Multiple reference signals collaborative sensing for integrated sensing and communication system towards {5G-A} and {6G}},
  author={Wei, Zhiqing and Li, Fengyun and Liu, Haotian and Chen, Xu and Wu, Huici and Han, Kaifeng and Feng, Zhiyong},
  journal={IEEE Transactions on Vehicular Technology},
  year={2024},
  publisher={IEEE}
}

@inproceedings{strinati2024distributed,
  title={Distributed intelligent integrated sensing and communications: The {6G-DISAC} approach},
  author={Strinati et al., Emilio Calvanese},
  booktitle={2024 Joint European Conference on Networks and Communications \& 6G Summit (EuCNC/6G Summit)},
  pages={392--397},
  year={2024},
  organization={IEEE}
}

@phdthesis{braun2014ofdm,
  title={{OFDM} radar algorithms in mobile communication networks},
  author={Braun, Klaus Martin},
  year={2014},
  school={Karlsruhe, Karlsruher Institut f{\"u}r Technologie (KIT), Diss.}
}

\end{document}